\documentclass[pra,tightenlines,superscriptaddress,twocolumn]{revtex4-1}

\usepackage{amsmath}
\usepackage{amssymb}
\usepackage{graphicx}
\usepackage{psfrag}
\usepackage{color}
\usepackage[utf8]{inputenc}
\usepackage{ifthen}
\usepackage{listings}
\usepackage{placeins}
\usepackage{mathrsfs}
\usepackage{autobreak}
\allowdisplaybreaks
\usepackage[margin=2cm]{geometry}
\usepackage{soul}

\usepackage[symbol]{footmisc}

\usepackage{mathrsfs}

\usepackage{dutchcal} 
\begin{document}

\title{Photon-assisted Landau Zener transitions in a tunable driven Rabi dimer coupled to a micromechanical resonator}

\author{Daniel Melvin}
\affiliation{Division of Materials Science, Nanyang Technological University, Singapore 639798, Singapore}

\author{Fulu Zheng}
\email{fzheng@uni-bremen.de}
\affiliation{Bremen Center for Computational Materials Science, University of Bremen, Bremen 28359, Germany}

\author{Kewei Sun}
\affiliation{School of Science, Hangzhou Dianzi University, Hangzhou 310018, China}

\author{Zhengjie Tan}
\affiliation{School of Mechanical and Aerospace Engineering, Nanyang Technological University, Singapore 639798, Singapore}

\author{Yang Zhao}
\email{yzhao@ntu.edu.sg}
\affiliation{Division of Materials Science, Nanyang Technological University, Singapore 639798, Singapore}

\begin{abstract}
Employing the multiple Davydov D$_2$ Ansatz with the time-dependent variational principle, we have investigated photon-assisted Landau-Zener (LZ) transitions and qubit manipulation in a hybrid quantum electrodynamics device. Modelled as a Rabi dimer, the device comprises of two interacting transmission-line resonators, each coupled to a qubit. The qubits, driven by independent harmonic fields, are further modulated by a micromechanical resonator mimicked by a phonon mode. The impacts of two independent driving fields on the qubit dynamics are carefully examined. The energy diagram of the system and the photon number mobilization on the resonators are analyzed to explain the behaviour of the LZ transitions and qubit dynamics while taking into account the influence of the single phonon mode. Results show that low phonon frequencies can alter the qubit dynamics, particularly in the absence of the driving fields, {and a strong phonon coupling strength can significantly perturb the qubit dynamics thanks to a high influx of phonon energy}. Notably, only the photon frequency affects the oscillation frequency of qubit polarization. This study unveils the imperative roles that photons and phonons play in the Rabi dimer model.
\end{abstract}
\date{\today}

\maketitle

\section{introduction}

The Landau-Zener-Stuckelberg-Majorana model \cite{landau_1932, zener_1932, stueckelberg_1932, majorana_1932}, also known as the Landau-Zener (LZ) model, describes a two-level system that is driven externally and has a time-dependent energy gap between its two diabatic states. As the energy separation switches signs, the diabatic states, which are the Hamiltonian eigenstates without tunneling, experience a level crossing. Transitions between the diabatic states are denoted as the LZ transitions. Conversely, the adiabatic states, which are the Hamiltonian eigenstates with tunneling, face an avoided crossing \cite{oleh_2022}. Characterizing both diabatic and adiabatic transitions, the LZ model has been widely adopted to investigate fundamental physical problems in various fields, such as atomic and molecular physics \cite{Zhu_NatureComm_2021}, condensed matter physics \cite{Lima_PRL_2010}, quantum information \cite{Cao_NatureComm_2013}, and quantum simulation \cite{Ivakhnenko_SciRep_2018}.

The LZ model can be formulated within the framework of circuit quantum electrodynamics (QED) systems, where two-level objects are typically coupled to resonators \cite{Wallraff_Nature_2004, Blais_PRA_2004}. Representing advanced QED apparatuses, hybrid circuit QED devices \cite{Houck_Nat_Phys_2012, Fitzpatrick_PRX_2017,
Kounalakis_QuanInf_2018, Landig_NatComm_2019, martinez_2019} are among the most promising candidates for realizing quantum information processing \cite{Devoret_Science_2013, atharv_2021} and quantum computation \cite{essammouni_2017, Forn_Diaz_Rev_Mod_Phys_2019, Arute_Nature_2019}. Investigating the LZ dynamics on such hybrid circuit QED platforms not only benefits fundamental studies of light-matter interactions, but also facilitates developments in quantum computation. 

In an advancement of a single-qubit QED model, hybrid QED devices, a system containing two coupled resonators each connected to one qubit could be an ideal model to unveil underlying physical mechanisms from both experimental and theoretical perspectives. The exceptional capability of hybrid QED models to characterize many-body quantum dynamics \cite{Kurcz_2014, you_2014, martinez_2019, Mahmoodian_2020} has attracted great interest  \cite{hu_2012, lambert_2018,contretas_2013, lambert_2013,viennot_2014, cubaynes_2020}. Such a minimalistic QED lattice has been fabricated using transmission-line resonators and transmon qubits \cite{raftery_2014}. Applying the rotating wave approximation (RWA) to the qubit-resonator interaction, one can use a pair of the Jaynes-Cummings Hamiltonian \cite{jaynes_1963} to model this system \cite{schmidt_2010,  larson_2021, shapourin_2016}. Nevertheless, the RWA may be invalid in the regime of ultra-strong qubit-resonator coupling \cite{forn_2019, mercurio_2022}. In contrast, the counter-rotating terms are considered in the Rabi model \cite{Rabi_PR_1936, Rabi_PR_1937}. Hwang {\it et al.} constructed a Rabi dimer model and studied the phase transition of photons in two coupled resonators \cite{hwang_2016}. The combined effects of qubit-photon coupling and photon tunnelling rate on the photon dynamics are reported. Recently, with the intention to mimic environmental modulations on the Rabi dimer, we have proposed to couple the qubits to micromechanical resonators and investigated the photon-qubit-phonon dynamics by explicitly treating all the degrees of freedom (DOFs) in the hybrid system \cite{zheng_2021, huang_2019, zheng_2018}. The LZ transitions in such systems have also been studied by applying an external harmonic driving field to one of the qubits \cite{zheng_2021}. Nonetheless, the dependence of the qubit-photon dynamics on two independently controlled driving fields and phonon frequencies remains obscure.

In this study, our objectives are to understand the impacts of driving field amplitude and phase on the dynamics of LZ transition in a Rabi dimer and to elucidate the impacts of a common phonon mode on the qubit and the photon dynamics. The state of the hybrid system is expressed with the Davydov D$_2$ {\it Ansatz} and derived using the time-dependent variational principle. By tuning the external driving field amplitudes and phases, the left and right qubit dynamics show distinct patterns of the LZ transitions. Additionally, the presence of the photon tunnelling allows the photons to hop between the left and right qubits. Thereafter, low frequency phonons and qubit-phonon coupling will be introduced to the Rabi Dimer system to emulate environmental effects in the Rabi dimer system.

The remainder of this paper is structured as follows Sec.~\ref{methodology} we present our methodology, including the system Hamiltonian, the Multi D$_2$ {\it Ansatz}, and the time-dependent variational principle. In Sec.~\ref{results}, results and discussions will be given. Finally, conclusions are drawn in Sec.~\ref{Conclusions}.

\section{methodology}
\label{methodology}
\subsection{Hamiltonian of the hybrid QED device}

\begin{figure}
  \includegraphics[scale=0.35]{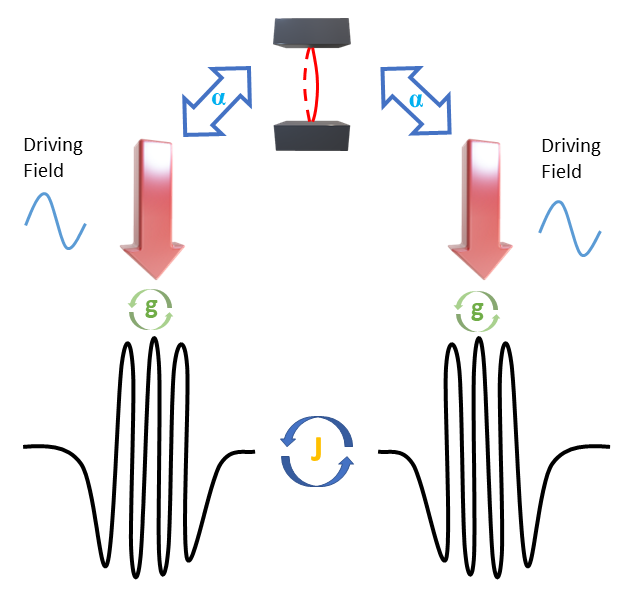}
  \caption{A sketch of the hybrid QED system under study. Photons can hop between two transmission-line resonators with a tunneling rate of $J$. Driven by an external field, the left (right) qubit is coupled to the photon mode in the left(right) resonator with a coupling strength $g$. Both left and right qubits are coupled to a common phonon mode with a coupling strength $\alpha$.}
  \label{Fig1_schem}
\end{figure}

As illustrated in Fig.~\ref{Fig1_schem}, the hybrid circuit QED system explored in the current work comprises of two coupled transmission-line resonators each interacting with a qubit, and the qubits are modulated via external driving fields and coupled to a micromechanical resonator. The two transmission-line resonators coupled with qubits are modelled by a Rabi dimer Hamiltonian ($\hbar=1$)
\begin{equation}\label{eq:HRD}
	H_{\textrm{RD}}=H^{\textrm{Rabi}}_{\textrm{L}}+H^{\textrm{Rabi}}_{\textrm{R}}-J(a_{\textrm{L}}^{\dagger}a_{\textrm{R}}+a_{\textrm{R}}^{\dagger}a_{\textrm{L}}),
\end{equation}
where $H_{\textrm{L}/\textrm{R}}^\textrm{Rabi}$ is a Rabi Hamiltonian describing the left (L) or right (R) resonator coupled to a qubit, and $J$ refers to the tunnelling rate for photons hopping between the two resonators. Specifically, the Rabi Hamiltonian,~\cite{Rabi1936, Rabi1937, braak2011, zhong2017}
\begin{equation}\label{Hrabi}
	H_{i=\textrm{L}, \textrm{R}}^\textrm{Rabi} = \frac{F_{i}}{2}\cos(\Omega_{i}t+\Phi_{i}) \sigma_{z}^{i} + \omega_{i} a_{i}^{\dagger} a_{i} - g_{i} ( a_{i}^{\dagger} + a_{i} ) \sigma_{x}^{i}, 
\end{equation}
presents a driven qubit coupled to a photon mode at frequency $\omega_i$ with a strength of $g_i$ in the $i$th ($i=\textrm{L}, \textrm{R}$) resonator. The external harmonic driving field imposed to the $i$th qubit is characterized by an amplitude $F_i$, a frequency $\Omega_{i}$ and an initial phase $\Phi_{i}$. Here $\sigma_{x}^{i}$ and $\sigma_{z}^{i}$ are Pauli matrices and $a_{i}$ ($a_{i}^{\dagger}$) is the annihilation (creation) operator of the $i$th photon mode. Throughout this work, it is assumed that the two photon modes have the same frequencies $\omega_{\textrm{L}}=\omega_{\textrm{R}}=\omega_{r}$, and the qubit-photon coupling strengths in two resonators are also the same, $g_{\textrm{L}}=g_{\textrm{R}}=g$.

The micromechanical resonator is modelled by a single phonon mode
\begin{equation}\label{Hb}
	H_\textrm{ph}= \omega_\textrm{ph} b^{\dagger} b
\end{equation}
with a frequency of $\omega_\textrm{ph}$ and a creation (annihilation) operator $b^{\dagger}$ ($b$). The interaction between the micromechanical resonator and the two qubits is then expressed as
\begin{equation}\label{Hbq}
	H_{\textrm{ph-q}}= \alpha (b^{\dagger}+b)(\sigma_{z}^{\textrm{L}}+\sigma_{z}^{\textrm{R}}),
\end{equation}
where $\alpha$ stands for the qubit-phonon coupling strength. Combining the Hamiltonians for the Rabi dimer, the phonon mode, and the qubit-phonon interaction yields the total Hamiltonian for the hybrid system
\begin{equation}\label{Htot}
	H=H_{\textrm{RD}}+H_\textrm{ph}+H_{\textrm{q-ph}}.
\end{equation}

Considering the low working temperatures of the modelled QED device, the micromechanical resonator in the system is thermally inactive and thus the temperature is set to zero in this work. Nevertheless, we still would like to comment that quantum dynamics at finite temperatures is also readily accessible within the undermentioned framework via the integration of methodologies, such as the Monte Carlo importance sampling method \cite{wang_2016}, the thermal field method \cite{chen_2017_thermal} and the method of displaced number states \cite{werther_2019, wangjun_2020}

\subsection{The multiple Davydov D$_2$ {\it Ansatz}}

To unveil the role of the phonon mode and driving fields in manipulating the photon and qubit states, we need a method which can explicitly depict all the DOFs for the qubits, the photon and phonon modes. Density matrix-based approaches are, therefore, inadequate for this purpose, as only the dynamics of the electronic DOFs are resolved after tracing out the bosonic DOFs. In contrast, the multiple Davydov D$_2$ (multi-D$_2$) {\it Ansatz} combined with the time-dependent variational principle treats all the DOFs on an equal footing and can capture the wave function propagation for all the DOFs\cite{zhao_2022,zhao_2023}. The outstanding performance of the multi D$_2$ {\it Ansatz} to deliver an exact solution to the Schrodinger equation even for complex many-body problems by using high multiplicity\cite{wang_2021,zhou_2014,zhou_2015,zhao_luo,zhao_1997,HuangZhao_2018,zhou_2018,lchen_2017,huang_fujihashi,huang_hoshina,huang_wang,jakucionis_2022,yan_2021,sun_2023}. Moreover, the approach has been successfully employed in simulating the quantum dynamics in similar hybrid QED systems with balanced numerical accuracy and efficiency. In this work, the multi-D$_2$ {\it Ansatz}
\begin{eqnarray}\label{eq:MD2}
	|{\rm D}_{2}^{M}(t)\rangle &=& \sum_{n=1}^{M} \Big[ A_{n} (t)|\uparrow\uparrow\rangle + B_{n} (t) |\uparrow\downarrow\rangle + C_{n} (t) |\downarrow\uparrow\rangle \nonumber\\
	&&~~~~+ D_{n} (t) |\downarrow\downarrow\rangle \Big] \bigotimes |\mu_{n}\rangle_{\textrm{L}}|\nu_{n}\rangle_{\textrm{R}} |\eta_{n}\rangle_{\textrm{ph}},
\end{eqnarray}
with a multiplicity $M$ is adopted to study the coupled photon-qubit-phonon dynamics in the QED device. Here $|\uparrow \downarrow \rangle=| \uparrow \rangle_{\textrm{L}} \otimes | \downarrow \rangle_{\textrm{R}}$ with $\uparrow$ $(\downarrow)$ representing the up (down) state of the qubits, and $|\mu_{n}\rangle_{\textrm{L}}$ ($|\nu_{n}\rangle_{\textrm{R}}$) and $|\eta_{n}\rangle_{\textrm{B}}$ are the coherent states for the left (right) photon mode and the phonon mode, respectively. The coherent states are
\begin{eqnarray}
	|\mu_{n}\rangle_{\textrm{L}} & = & \exp\left[\mu_{n} (t) a_{\textrm{L}}^{\dagger}-\mu_{n}^{\ast} (t) a_{\textrm{L}}\right]|0\rangle_{\textrm{L}},\\
	|\nu_{n}\rangle_{\textrm{R}} & = & \exp\left[\nu_{n} (t) a_{\textrm{R}}^{\dagger}-\nu_{n}^{\ast} (t) a_{\textrm{R}}\right]|0\rangle_{\textrm{R}}, \\
  |\eta_{n}\rangle_{\textrm{ph}} &=& \exp \left[ \eta_{n} (t)  b^{\dagger}-\eta_{n}^{\ast} (t) b \right] |0\rangle_{\textrm{ph}},
\end{eqnarray}
where $|0\rangle_{\textrm{L}(\textrm{R})}$ and $|0\rangle_{\textrm{B}}$ are the vacuum state of the left (right) photon mode and the phonon mode, respectively. The variational parameters $A_{n}(t)$, $B_{n}(t)$, $C_{n}(t)$, and $D_{n}(t)$ represent the probability amplitudes of corresponding qubit states. In addition, $\mu_{n}$ ($\nu_{n}$) and $\eta_{n}$ are the displacements of the left (right) photon mode and the phonon mode, respectively. These time-dependent variational parameters can be evaluated via the time-dependent variational principle, producing the wave function propagating along time.

\subsection{The time-dependent variational principle}

The equations of motion for all the variational parameters are derived from
\begin{equation}\label{DiracFrenkel}
 	\frac{d}{dt} \bigg( \frac{\partial \mathscr{L}}{ \partial \dot{\alpha}^{*}_{n}} \bigg) - \frac{\partial \mathscr{L}}{ \partial \alpha^{*}_{n}} =0,
\end{equation}
where $\alpha_{n}$ represents one of the aforementioned variational parameters, and $\mathscr{L}$ is the Lagrangian
\begin{equation}\label{Lagrangian}
	\mathscr{L} = \frac{i}{2} \langle {\rm D}_{2}^{M}(t) | \frac{\overrightarrow{\partial}}{\partial t} - \frac{\overleftarrow{\partial}}{\partial t} | {\rm D}_{2}^{M}(t) \rangle - \langle {\rm D}_{2}^{M}(t) | H | {\rm D}_{2}^{M}(t) \rangle.
\end{equation}
The derivation details can be found in Appendix~\ref{Equations of Motion}.

\subsection{Physical observables of interest}

As stated above, by combining the time-dependent variational principle with the multi-D$_2$ {\it Ansatz}, we can obtain the accurate time-dependent wave function of the entire system, which makes it straightforward to evaluate all the physical observables to characterize the coupled photon-qubit-phonon dynamics. The photon dynamics is presented by the time evolution of the photon population in two resonators
\begin{eqnarray}
	N_{\textrm{L}}(t) & = & \langle{\rm D}_{2}^{M}(t)| a_{\textrm{L}}^{\dagger} a_{\textrm{L}} |{\rm D}_{2}^{M}(t) \rangle \nonumber\\
	& = & \sum_{l,n}^{M} \Big[ A_{l}^{\ast}(t) A_{n}(t)  + B_{l}^{\ast}(t) B_{n}(t)  + C_{l}^{\ast}(t) C_{n}(t) \nonumber\\
	&&~~~~~~~ + D_{l}^{\ast}(t) D_{n}(t) \Big] \mu_{l}^{\ast}(t) \mu_{n}(t)  S_{ln}(t) , \\
	N_{\textrm{R}}(t) & = & \langle{\rm D}_{2}^{M}(t)|a_{\textrm{R}}^{\dagger}a_{\textrm{R}}|{\rm D}_{2}^{M}(t) \rangle \nonumber \\
	& = & \sum_{l,n}^{M} \Big[ A_{l}^{\ast}(t) A_{n}(t)  + B_{l}^{\ast}(t) B_{n}(t)  + C_{l}^{\ast}(t) C_{n}(t) \nonumber\\
	&&~~~~~~~ + D_{l}^{\ast}(t) D_{n}(t) \Big] \nu_{l}^{\ast}(t) \nu_{n}(t)  S_{ln}(t) ,
\end{eqnarray}
where $S_{ln} (t)$ is the Debye-Waller factor
\begin{eqnarray}
	S_{ln} & = &  \exp\left[\mu_{l}^{\ast}(t) \mu_{n}(t)-\frac{1}{2} |\mu_{l}(t)|^2-\frac{1}{2}|\mu_{n}(t)|^2\right] \cdot \nonumber\\
	&& \exp\left[ \nu_{l}^{\ast}(t) \nu_{n}(t) - \frac{1}{2} |\nu_{l}(t)|^2 - \frac{1}{2} |\nu_{n}(t)|^2 \right] \cdot \nonumber\\
	&&\exp \left[ \eta_{l}^{\ast}(t) \eta_{n}(t) - \frac{1}{2} |\eta_{l}(t)|^2 - \frac{1}{2} |\eta_{n}(t)|^2 \right].  \nonumber
\end{eqnarray}
With $N_{\textrm{L}}(t)$ and $N_{\textrm{R}}(t)$, the time evolution of the total photon number ${N}(t)=N_{\textrm{L}}(t)+N_{\textrm{R}}(t)$ and photon imbalance ${Z}(t)=N_{\textrm{L}}(t)-N_{\textrm{R}}(t)$ can be directly obtained.

In addition to the photon dynamics, the time evolution of the qubit states is recorded during the simulations by measuring the time evolution of the qubit polarization
\begin{eqnarray}
	\langle\sigma_{z}^{\textrm{L}}(t)\rangle & = & \langle{\rm D}_{2}^{M}(t)|\sigma_{z}^{\textrm{L}}|{\rm D}_{2}^{M}(t) \rangle \nonumber \\
	& = & \sum_{l,n}^{M} \Big[ A_{l}^{\ast}(t)A_{n}(t) + B_{l}^{\ast}(t)B_{n}(t) \nonumber\\
	&&- C_{l}^{\ast}(t)C_{n}(t) - D_{l}^{\ast}(t)D_{n}(t) \Big] S_{ln}(t),\\
	\langle\sigma_{z}^{\textrm{R}}(t)\rangle & = & \langle{\rm D}_{2}^{M}(t)|\sigma_{z}^{\textrm{R}}|{\rm D}_{2}^{M}(t) \rangle \nonumber \\
	& = & \sum_{l,n}^{M} \Big[ A_{l}^{\ast}(t)A_{n}(t) - B_{l}^{\ast}(t)B_{n}(t) \nonumber\\
	&&+ C_{l}^{\ast}(t)C_{n}(t) - D_{l}^{\ast}(t)D_{n}(t) \Big] S_{ln}(t).
\end{eqnarray}

The LZ transition probabilities, quantifying the qubit flipping probabilities from the down states to the up states, are expressed as
\begin{eqnarray}
	\ P_{\textrm{LZ}}^{\textrm{L}}(t) & = & |\langle\uparrow_{\textrm{L}}|{\rm D}_{2}^{M}(t) \rangle|^2 \nonumber \\
	& = & \sum_{l,n}^{M} \Big[ A_{l}^{\ast}(t)A_{n}(t) + B_{l}^{\ast}(t)B_{n}(t) \Big] S_{ln}(t)\\
	\ P_{\textrm{LZ}}^{\textrm{R}}(t) & = & |\langle\uparrow_{\textrm{R}}|{\rm D}_{2}^{M}(t) \rangle|^2 \nonumber \\
	& = & \sum_{l,n}^{M} \Big[ A_{l}^{\ast}(t)A_{n}(t) + C_{l}^{\ast}(t)C_{n}(t) \Big] S_{ln}(t).
\end{eqnarray}

Similar to the photon number, the phonon population can be monitored via
\begin{eqnarray}
	N_\textrm{ph}(t) &=& \langle{\rm D}_{2}^{M}(t)| b^{\dagger} b|{\rm D}_{2}^{M}(t) \rangle \nonumber \\
	&=& \sum_{l,n}^{M} \Big[ A_{l}^{\ast}(t) A_{n}(t) + B_{l}^{\ast}(t) B_{n}(t) + C_{l}^{\ast}(t) C_{n}(t) \nonumber\\
	&&~~~~~~~~+ D_{l}^{\ast}(t) D_{n}(t) \Big] \eta_{l}^{\ast}(t) \eta_{n}(t) S_{ln}(t).
\end{eqnarray}
and the phonon energy $E_\textrm{ph}(t) = \omega_\textrm{ph} N_\textrm{ph}(t)$ is obtained in turn.

\subsection{Parameter configurations and initial conditions}

The photon-qubit coupling strength $g$ and the photon hopping rate $J$ significantly impact the qubit dynamics in our QED system. Hwang {\it et al.} found a critical photon tunnelling rate $J_{c}=0.03~\omega_{0}$ in a bare resonant the Rabi dimer (qubit bias is resonant to the photon frequency $\omega_{r}$)~\cite{hwang_2016}, above which the photons are delocalized over two resonators regardless of the photon-qubit coupling strength. Here $\omega_{0}$ stands for the energy unit in this study. An ultra-strong photon-qubit coupling strength of $g=0.3~\omega_{0}$ is adopted in the current work. In order to unveil the effects of photon population on the driven qubits, we consider two photon-tunnelling rates $J=0.01~\omega_{0}$ and $J=0.075~\omega_{0}$, which lead to photon localization and delocalization, respectively, in a bare resonant Rabi dimer~\cite{hwang_2016, zheng_2018}.

The left and right photon modes have the same frequency $\omega_\textrm{L}=\omega_\textrm{R}=\omega_{r} =10~\omega_{0}$. The high photon frequency makes the relaxation time of each LZ transition shorter than the interval between two neighbouring LZ transitions \cite{zheng_2021}. Initially, twenty photons $N_{\textrm{L}}(0)=20$ are pumped into the left resonator, while the right resonator is at vacuum state with $\mu_1(t=0)=\sqrt{20}$ and $\mu_{n\neq1}(t=0)=\nu_n(t=0)=0$. Both the left and right qubits are prepared in the down state with $A_n(t=0)=B_n(t=0)=C_n(t=0)=D_{n\neq1}(t=0)=0$ and $D_1(t=0)=1$. The single phonon mode is in the vacuum state $\eta_{n}(t = 0) = 0$ initially. Based on extensive convergence tests with respect to a multiplicity of $M = 6$ is adopted in this study, which provides converted results with excellent efficiency.

\section{Results and discussion}\label{results}

Our study focuses on manipulating qubit states via adjusting external driving fields and the coupling between qubits and a micromechanical resonator ($g$). In Sec.~\ref{sub_sec_field}, we examine the impact of the driving fields on the qubit-photon dynamics in a Rabi dimer. In Sec.~\ref{sub_sec_phonon}, the qubit-phonon coupling is activated, and its effect on the photon-qubit-phonon dynamics is examined.

\subsection{Manipulating LZ transitions in a Rabi dimer through tuning the external driving field}\label{sub_sec_field}

In Hamiltonian (\ref{Hrabi}), two independent harmonic driving fields are applied to two qubits. This allows for direct control of the qubit states \cite{shi_2021}. We consider the two driving fields having the same frequency of ($\Omega_{\textrm{L}}=\Omega_{\textrm{R}}=0.05~\omega_{0}$), but with varying amplitudes ($F_\textrm{i}$) and initial phases ($\phi_\textrm{i}$). The influence of these driving fields on the dynamics of the qubit-photon system is analyzed in the subsequent two subsections.

\subsubsection{Impact of driving field amplitude on LZ transitions}\label{sub_sub_sec_field_amplitudes}

\begin{figure*}[!ht]
  \centering
  \includegraphics[scale=0.5]{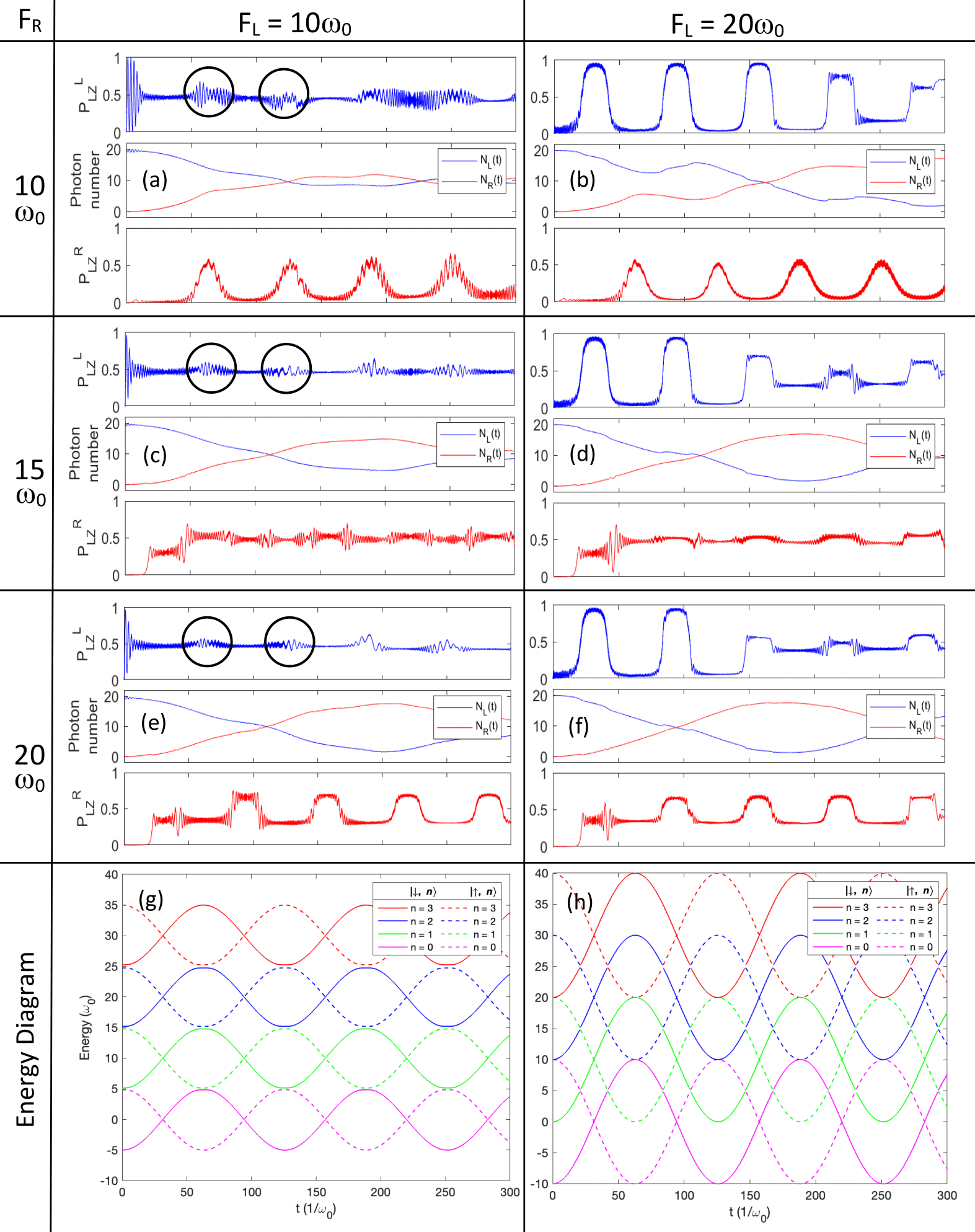}
  \caption{ Dynamics of the Rabi dimer system and energy diagrams with the initial phase of the driving field $\Phi_{\textrm{L}} = \Phi_{\textrm{R}} = 0$ and photon frequency $\omega_\textrm{L}=\omega_\textrm{R}=\omega_{r} =10~\omega_{0}$. The panels have different driving field strengths parameterised in the following way; 
  a) $F_{\textrm{L}}=10~\omega_{0}$ and $F_{\textrm{R}}=10~\omega_{0}$; 
  b) $F_{\textrm{L}}=20~\omega_{0}$ and $F_{\textrm{R}}=10~\omega_{0}$; 
  c) $F_{\textrm{L}}=10~\omega_{0}$ and $F_{\textrm{R}}=15~\omega_{0}$; 
  d) $F_{\textrm{L}}= 20~\omega_{0}$ and $F_{\textrm{R}}=15~\omega_{0}$; 
  e) $F_{\textrm{L}}=15~\omega_{0}$ and $F_{\textrm{R}}=20~\omega_{0}$; 
  f) $F_{\textrm{L}}= 20~\omega_{0}$ and $F_{\textrm{R}}=20~\omega_{0}$; 
  g) Energy diagram of a qubit with $F_{\textrm{L}, \textrm{R}} = 10~\omega_{0}$; 
  h) Energy diagram of a qubit with $F_{\textrm{L}, \textrm{R}} = 20~\omega_{0}$. An avoided crossing happens whenever neighbouring number states intersect one another. The energy diagrams only show the left qubit, and the photon number states $n$ = 0, 1, 2, and 3 even though the number states could be infinite.}
  \label{Fig2}
\end{figure*}

In order to highlight the effects of the driving field amplitude on the coupled qubit-photon dynamics in the Rabi dimer, the two fields are first configured with an identical initial phase $\Phi_{\textrm{L}} = \Phi_{\textrm{R}} = 0$, but with varying amplitudes. In a bare resonant Rabi dimer ($\lambda = 0$) given by Hamiltonian (\ref{eq:HRD}) with a photon tunnelling rate $J=0.01~\omega_0$ and a qubit-photon coupling strength $g = 0.3~\omega_0$, the photons initially pumped into the left resonator will be localized in the resonator in the absence of the driving fields. Once the external fields are imposed on the qubits, the photon distribution is dramatically affected, as shown in Fig.~\ref{Fig2}, where the driving field amplitudes of $F_{\textrm{L}} \in \{10~\omega_{0}, 20~\omega_{0}\}$ and $F_{\textrm{R}} \in \{10~\omega_{0}, 15~\omega_{0}, 20~\omega_{0}\}$ are configured to show their effects on the qubit-photon dynamics.

The results for $F_{\textrm{L}} = 10~\omega_{0}$ and $20~\omega_{0}$ are presented in the left (panels a, c, and e) and right (panels b, d, and f) columns of Fig.~\ref{Fig2}, respectively. The three panels in each column contain the qubit-photon dynamics from different amplitudes of the right driving field. One can immediately find that the dynamics of the left (right) qubit are dominated by the left (right) driving field by comparing the results in each row (column) of Fig.~\ref{Fig2}. For instance, in Fig.~\ref{Fig2} (a, c and e), the left driving field amplitude $F_{\textrm{L}}=10~\omega_{0}$ leads to similar LZ transition patterns in the left qubit. There is a damped oscillation in $P_{\textrm{LZ}}^{\textrm{L}}$ in the short time after $\omega_0 t=0$, as the hybrid states $|\downarrow, n \rangle$ and $|\uparrow, n-1\rangle$ intersect right at the start ($\omega_0 t\approx 0$) with $F_{\textrm{L}} = 10~\omega_{0}$ (see the energy diagram in Fig.~\ref{Fig2}(g)). Here, $|\downarrow (\uparrow) \rangle$ is the left qubit up (down) state, and $| n \rangle$ is the number state of the left photon mode. Both the photons initially localized in the left resonator and the vanishing facilitate energy gap between the above states at $\omega_0 t \approx 0$ the flipping of the left qubit. Then, as the above hybrid states separate in energy, $P_{\textrm{LZ}}^{\textrm{L}}$ approaches $0.5$ with weak oscillations. In comparison to the frequent flipping of the left qubit, the right qubit remains on its initial down state due to the low photon population in the right resonator at $\omega_0 t \approx 0$. With a driving amplitude of $F_{\textrm{R}} = 10~\omega_{0}$, there is no avoided crossing, as can be seen from the energy diagram in Fig.~\ref{Fig2}(g). Therefore, there are no sudden changes on the right qubit dynamics (see Fig.~\ref{Fig2} (a) and (b)). Instead, the right qubit exhibits a gradual population transfer to its up state as two neighboring adiabatic states approach each other in energy. As the two states separate in energy, the right qubit repopulates in the down state, resulting in the pulses in $P_{\textrm{LZ}}^{\textrm{R}}$ shown in Fig.~\ref{Fig2} (a) and (b). During the intersection between two adjacent adiabatic states, $P_{\textrm{LZ}}^{\textrm{R}}$ is at the peak of the pulse, and $P_{\textrm{LZ}}^{\textrm{L}}$ also shows pronounced oscillations as highlighted by the black circles in Fig.~\ref{Fig2}. 

In contrast to the dynamics of $F_{\textrm{L}} = 10~\omega_{0}$, the left qubit driven by a field of $F_{\textrm{L}} = 20~\omega_{0}$ is localized in its down state from the beginning until it meets the first avoided crossing point, as illustrated in Fig.~\ref{Fig2} (b), (d) and (f). The detuning between $|\downarrow, n \rangle$ and $|\uparrow, n-1\rangle$ at $\omega_0 t \approx 0$ hinders the flipping of the left qubit. With the assistance from sufficient photons in the left resonator, the left qubit can realize complete population inversion between its up and down states at the first few avoided crossings, leading to the square-waved patterns of $P_{\textrm{LZ}}^{\textrm{L}}$ in Fig.~\ref{Fig2} (b), (d) and (f). The hybrid qubit-photon subsystem in the left resonator evolves adiabatically through these avoided crossings long paths of $|\downarrow, n+2\rangle \rightarrow |\uparrow, n+1\rangle \rightarrow |\downarrow, n\rangle \rightarrow |\uparrow, n+1\rangle \cdots$. The energy diagram in Fig.~\ref{Fig2} (h) provides a straightforward way to analyze such transition paths composed of diabatic qubit-photon states.

\begin{figure*}[!ht]
  \centering
  \includegraphics[scale=0.5]{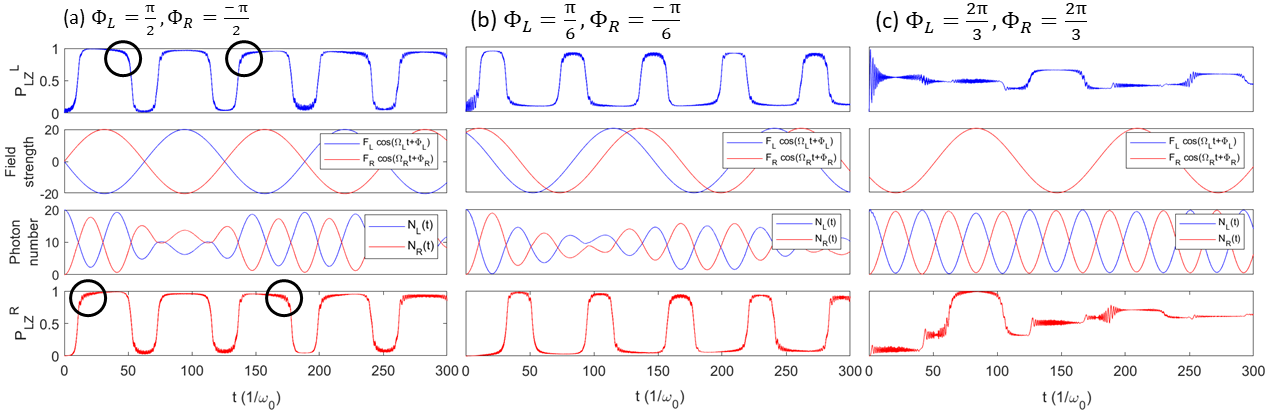}
  \caption{The dynamics of the Rabi dimer system in the presence of driving fields on both qubits with $F_{\textrm{L}} = F_{\textrm{R}} = 20~\omega_{0}$ and photon frequency $\omega_\textrm{L}=\omega_\textrm{R}=\omega_{r} =10~\omega_{0}$. The panels are characterised by the different initial phases of the driving fields: a) $\Phi_{\textrm{L}}=- \Phi_{\textrm{R}}=\pi/2$, b) $\Phi_{\textrm{L}}=- \Phi_{\textrm{R}}=\pi/6$, and c) $\Phi_{\textrm{L}}=\Phi_{\textrm{R}}=2\pi/3$.}
  \label{Fig3}
\end{figure*}

It is noteworthy that the sudden flipping of the left qubit between its up and down states with $F_{\textrm{L}} = 20~\omega_{0}$ is also known as the adiabatic rapid passage (ARP) \cite{camparo_1984}, which is typically adopted to achieve efficient population transfer in driven quantum systems \cite{Chanda_2023_ARP}. As the name suggests, a successful ARP only occurs when three conditions are met. Firstly, the process has to be adiabatic, i.e., field-induced sweep on the detuning is sufficiently slow compared to the period of resonant Rabi oscillation. Secondly, the sweep should be more rapid than the excited state relaxation. This condition is always fulfilled in this section as the system is operating at zero temperature. Thirdly, at the endpoints (both the starting and ending points), field-induced detuning must be far from resonance to the energy gap between the adiabatic states of the quantum system. 

The first condition for ARP requires that $|\upsilon|/\Delta \ll \Delta$ \cite{camparo_1984, Chanda_2023_ARP}. Here the sweep rate $\upsilon$ is estimated using the time derivative of the driving field $\upsilon = -\Omega_i F_{i}\sin(\Omega_{i}t+\Phi_{i})/2$ at avoided crossings and $\Delta=2g\sqrt{n+1}$ is approximately the energy gap between the adiabatic states \cite{kenmoe_2017,vitanox_1999,Sun_LZ_PRA_2012}. For the hybrid qubit-photon subsystem in the left resonator with $F_{\textrm{L}} = 20~\omega_{0}$ (see Fig.~\ref{Fig2} (b), (d) and (f)), this condition is satisfied at all avoided crossings with $|\upsilon|= \frac{\sqrt{3}}{4} \omega_{0}^2$ and $\Delta^2 = 0.36 (n+1) \omega_{0}^2$, as states $|\downarrow, n\rangle$ with $n \geq 1$ will be populated initially. However, this adiabaticity criterion cannot be fulfilled in the right subsystem due to insufficient initial photons even if the driving field has an amplitude of $F_{\textrm{R}} = 20~\omega_{0}$ (see Fig.~\ref{Fig2} (e) and (f)). Instead of a successful ARP with $100\%$ population transferred to the up state, the right qubit exhibits a LZ at first avoided crossing with $\sim 40\%$ of its population adiabatically flows to the up state, as illustrated in Fig.~\ref{Fig2} (e) and (f).

The third condition for ARP ensures that the system is localized in an adiabatic state before and after the ARP. As aforementioned and illustrated in the right column of Fig.~\ref{Fig2}, the initial large detuning between $|\downarrow, n \rangle$ and $|\uparrow, n-1\rangle$ induced by $F_{\textrm{L}} = 20~\omega_{0}$ confines the left qubit on its initial down state before the first avoided crossing. Therefore, multiple ARPs can be realized on the left qubit with all three conditions satisfied. Tuning the driving field amplitude to $F_{\textrm{L}} = 10~\omega_{0}$ leads to close energies for the above adiabatic states (see  Fig.~\ref{Fig2} (g)) and left qubit delocalization at $\omega_0 t \approx 0$, hindering ARP on the left qubit. In addition to changing the driving field amplitudes, initial detuning between the adiabatic states and resulting ARP can be manipulated by tuning the initial phases of the driving fields, as demonstrated in the following section.

\subsubsection{Effects of the initial driving field phase on LZ transitions}\label{sub_sub_sec_field_phase}

Besides the driving field amplitude, the initial phase of the driving field is another critical parameter that directly impacts the energy diagram of qubit-photon coupling and the system dynamics. In this section, a photon tunneling rate $J = 0.075~\omega_{0}$ is configured, which allows the photons to hop between the two resonators even with a strong qubit-photon coupling of $g=0.3~\omega_{0}$. The amplitude and frequency of the driving fields are set to $F_{\textrm{L}} = F_{\textrm{R}} = 20~\omega_{0}$ and  $\Omega_\textrm{L}=\Omega_\textrm{R}= 0.05~\omega_{0}$, respectively. Therefore, delocalized photons over two resonators ensure that the adiabaticity criterion for ARP is fulfilled. Tuning the initial phase of the driving fields can allow control over the system dynamics, as presented in Fig.~\ref{Fig3} with several examples. 

With $\phi_{\textrm{L}} = -\phi_{\textrm{R}} = \pi/2$, the two fields lead to avoided crossings at the same times for the left and right subsystems. Hence, the two qubits exhibit similar dynamics containing a sequence of ARPs, as shown in Fig.~\ref{Fig3} (a). It is detectable that there are some high-frequency oscillations right before or after the ARPs occur, as highlighted by the circles in Fig.~\ref{Fig3} (a). These oscillations appear at different time points for the two qubits. Comparing the qubit dynamics with the photon number evolution, one can find that these oscillations for a qubit only arise when the photon number in the corresponding resonator arrives at a local maximum. This phenomenon results from stronger effective tunneling as the photon population is increased (see Eq.(\ref{Hrabi})), thus leading to pronounced oscillations in $P_{\textrm{LZ}}$.  

The synchronization of two qubits shown in Fig.~\ref{Fig3} (a) can be readily removed by parameterizing the two fields with a different initial phase. For instance, presented in Fig.~\ref{Fig3} (b) is the situation with $\phi_{\textrm{L}} = -\phi_{\textrm{R}} = \pi/6$. Avoided crossings of the left subsystem do not occur at the same time as the right one. Therefore, the occurrence of the ARPs on the right qubit is delayed as compared to the left qubit. It can be clearly seen from Fig.~\ref{Fig3} (a) and (b) that diverse ARPs on the qubits can be achieved by adjusting the driving field phases. 

It can also happen that a specific initial phase configuration leads to neighboring adiabatic states having a small energy gap at $\omega_0 t=0$. Then the qubit will be immediately depolarized ($P_{\textrm{LZ}} \rightarrow 0.5$) after several flipping with the assistance from sufficient photons, such as the left qubit shown in Fig.~\ref{Fig3} (c) with $\phi_{\textrm{L}} = \phi_{\textrm{R}} = 2\pi/3$. Therefore, no ARP will take place at any of the avoided crossings. The qubits then tend to change their states at avoided crossings with the help of photons. This scenario is predominant in the dynamics of the right qubit. At the first avoided crossing, a LZ transition occurs with $\sim 30\%$ population transferred to its up state due to a small photon number in the right resonator. As the second avoided crossing arrives, almost all the photons are accumulated in the right resonator, which completely pumps the right qubit to its up state. 

As discussed above, both the driving field amplitude and initial phase greatly influence the coupled qubit-photon dynamics. Photon-assisted LZ transitions will take place at avoided crossings. Specifically, a sequence of ARPs can occur if the adiabaticity criterion is satisfied and the initial gap between the adjacent states $|\downarrow, n \rangle$ and $|\uparrow, n\pm1\rangle$ is large. The state evolution paths of the hybrid qubit-photon system can be uncovered by analyzing the qubit-photon dynamics with the energy diagram.

\subsection{Combined field and phonon effects on the Rabi dimer}\label{sub_sec_phonon}

Beyond the driving field effects on the LZ transitions of the Rabi dimer, environmental modulations on the hybrid qubit-photon dynamics are further investigated in this subsection. As illustrated in Fig.~\ref{Fig1_schem}, a phonon mode ($H_{\textrm{ph}}$) is coupled to the qubits via the interaction Hamiltonian ($H_{\textrm{q-ph}}$), producing the total Hamiltonian ($H$) in Eq.~(\ref{Htot}) for the composite system.

In order to understand the phononic perturbations in the Rabi dimer system, prior knowledge of the coupled qubit-photon dynamics in the absence of the phonon mode is required. In a resonant Rabi dimer system with $g = 0.3~\omega_0$ and $J = 0.01~\omega_0$, it has been discovered that photons are trapped in the initial resonator, i.e., in the localized phase due to the frequent qubit flipping~\cite{hwang_2016, zheng_2018}. In contrast, high-frequency photons with $\omega_r = 10~\omega_0$ will be delocalized over two resonators as qubit flipping rarely occurs~\cite{zheng_2021}. In the presence of environmental modulation, Zheng {\it et al.} discovered that stronger qubit-phonon coupling causes $P_{\textrm{LZ}}$ to approach 0.5 in a shorter time~\cite{zheng_2021}. It is also highlighted that the phonon mode population depends on its frequency and the strength of its coupling to the qubits. Combined effects from the driving field and the phonon mode on the coupled phonon-qubit-photon dynamics are addressed here by tuning the phonon frequency and the qubit-phonon coupling strength. A parameter configuration of $F_{\textrm{R}} = 0$, $F_{\textrm{L}} = 20~\omega_{0}$, $\phi_{\textrm{L}}=0$, $\Omega_{\textrm{L}} = 0.05~\omega_{0}$, $g = 0.3~\omega_{0}$ and $J = 0.01~\omega_{0}$ is adopted in calculations to follow unless otherwise specified. A driving field is only applied to the left qubit to distinguish the field effects from the ones of phonons on the LZ transitions. 

\begin{figure*}[htbp]
  \centering
  \includegraphics[scale=0.545]{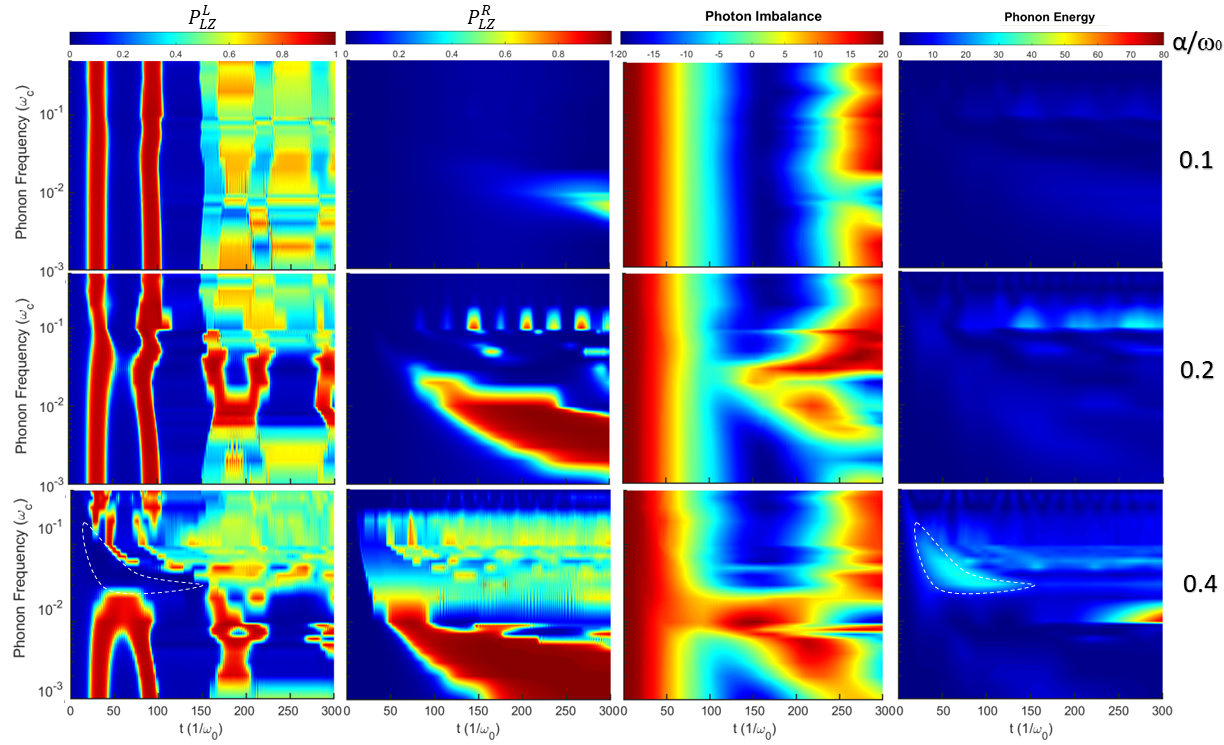}
  \caption{Time evolution of the LZ transition probability $P_{\textrm{LZ}}$ of the left (column 1) and right (column 2) qubits, photon imbalance (column 3) and phonon energy (column 4). A phonon mode with a frequency $\omega_\textrm{ph}$ in $0.001~\omega_{0}<\omega_\textrm{ph}<0.5~\omega_{0}$ is coupled to the qubits. The qubit-phonon coupling strength is set as $\alpha/\omega_0 = 0.1, 0.2$ and $0.4$, with the results for a $\alpha$ collected in an individual row of the figure. The driving field is only applied on the left qubit with $F_{\textrm{L}} = 20~\omega_{0}$ and $\Phi_{\textrm{L}} = 0$. The photon frequency is $\omega_{r} =10~\omega_{0}$ for all cases.}
  \label{Fig4}
\end{figure*}

A low-frequency phonon mode in the frequency range of $0.001~\omega_{0} \leq \omega_\textrm{ph} \leq 0.5~\omega_{0}$ is adopted to mimic the experimental micromechanic resonator coupled to the qubits. As illustrated in Fig.~\ref{Fig4}, three qubit-phonon coupling strengths, $\alpha/\omega_0 = 0.1, 0.2$ and $0.4$, are used to highlight the phonon mode effects on the system dynamics. 

With a weak qubit-phonon coupling strength of $\alpha = 0.1~\omega_0$, the phonon effects on Rabi dimer dynamics are negligible, and the qubit-photon dynamics shown in the upper row of Fig.~\ref{Fig4} is very similar to the case in the absence of the phonons \cite{zheng_2021}. The photons are delocalized over two resonators, and the right qubit is suppressed in its down state due to the large detuning between the right qubit and the photons. As time evolves, the phonon mode is gradually populated with the energy from the left qubit, which will effectively modulate the bias of the right qubit via the qubit-phonon coupling. If the phonon frequency is lower than $0.01~\omega_{0}$, noticeable phonon effects on the right qubit dynamics can be seen after $t\omega_{0} \approx 200$, as a phonon mode with lower frequency is easier to be populated. In this regime, the right qubit attempts to flip to its up state with the help of the photons in the right resonator, which in turn traps the photons in the right resonator. The phonon mode also modulates the bias of the left qubit, but such a modulation is weak compared to the driving field. Therefore, the phonon mode effects on the left qubit dynamics are negligible with a weak qubit-phonon coupling strength of $\alpha = 0.1~\omega_0$.

The phonon effects on the left qubit dynamics are more pronounced with $\alpha = 0.2~\omega_0$, as shown in the left panel of the middle row of Fig.~\ref{Fig4}. The energy flowing from the left qubit to the phonon mode is accelerated by increasing $\alpha$. Therefore, even the earlier ARPs in the left qubit for $t\omega_{0} < 100$ can feel the phonon modulation. With this stronger qubit-phonon coupling, more significant phonon effects on the right qubit dynamics are also seen from the advanced raising of $P_{\textrm{LZ}}^{\textrm{R}}$ compared to the situation for $\alpha = 0.1~\omega_0$.  The right qubit can flip to its up state and stay there for quite a long time if the phonon frequency is lower than $0.01~\omega_{0}$. The photon delocalization is also affected by the phonon-modulated qubit dynamics \cite{zheng_2018}. Increasing the qubit-phonon coupling strength further to $\alpha = 0.4~\omega_{0}$ also spotlights the impacts of the phonon mode on the coupled qubit-photon dynamics in the Rabi dimer, as illustrated in the lower row of Fig.~\ref{Fig4}. Compared to the results from weaker $\alpha$ shown in the top two rows of Fig.~\ref{Fig4}, the four sets of ARPs of the left qubit in $\omega_0 t<100$ are destroyed as the phonon energy increases tremendously with $\alpha = 0.4~\omega_{0}$. 

The key to understanding the significance of the phonon frequency in relation to qubit dynamics is by examining the population of the single phonon mode. A phonon mode with a low frequency facilitates the growth of the phonon population. However, it requires a long time to accumulate sufficient energy to affect the qubit dynamics. Interacting with both left and right qubits, the phonon mode bridges the energy flow between the two qubits. The left qubit distributes the energy from the driving field to both the phonon and the left photon mode via diagonal and off-diagonal coupling, respectively. The energy exchanges between the left qubit and the left photon mode accompanies the qubit flipping. This mechanism can be clearly seen by comparing the dynamics of the left qubit and the photon. For instance, with $\alpha = 0.1~\omega_0$ (cf. the first row of Fig.~\ref{Fig4}), the first four sets of ARPs of the left qubit occur with sudden changes in the photon imbalance. In contrast, qubit polarization (localized on one state) will stimulate the energy flow between the qubit and the phonon. This phenomenon is predominant when the qubit-phonon coupling is strong, such as the region highlighted by the dashed circles in the bottom row of Fig.~\ref{Fig4} with $\alpha = 0.4~\omega_0$.

\begin{figure}
  \centering
  \includegraphics[scale=0.4]{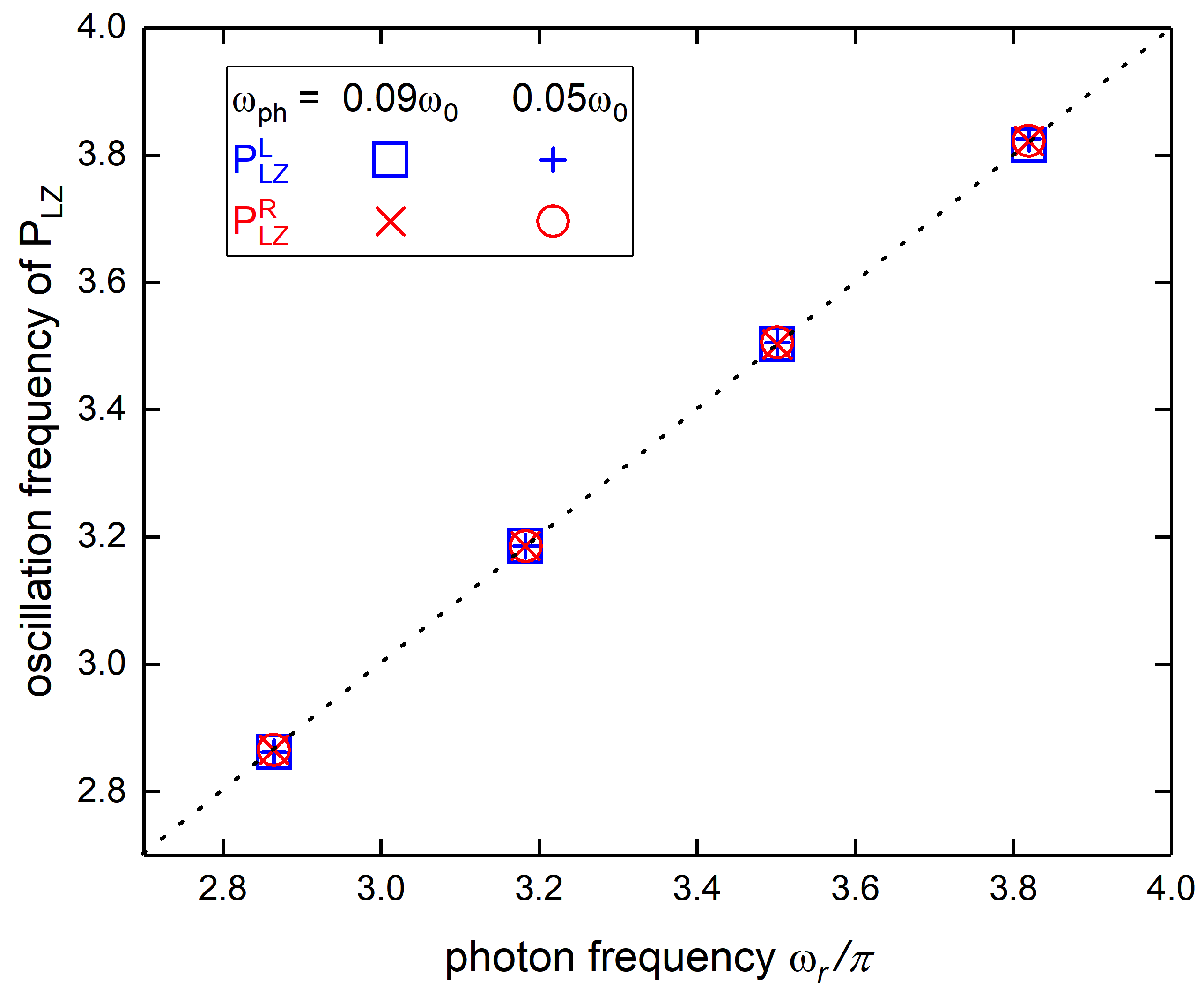}
  \caption{Oscillation frequency of the LZ transition probability $P_{\textrm{LZ}}$ versus the photon frequency. Calculations were performed with two different phonon frequencies ($\omega_\textrm{ph} = 0.05~\omega_{0}$ and $0.09~\omega_{0}$). The dashed line is a linear fitting.}\label{Fig5}
\end{figure}

It is noteworthy that the LZ transition probability $P_{\textrm{LZ}}$ exhibits persistent high-frequency oscillations with small amplitudes regardless of the presence or absence of phonon mode. In order to gain deeper insight into these oscillations, a Fourier transformation technique is employed to unveil the frequencies of these oscillations in $P_{\textrm{LZ}}$. Calculations with various frequencies for the photon and phonon modes are performed with $\omega_{r}/\omega_{0} \in \{9, 10, 11, 12\}$ and $\omega_{\textrm{ph}}/\omega_{0} \in \{0.05, 0.09\}$. The oscillation frequencies of $P_{\textrm{LZ}}$ from the calculations are plotted versus the photon frequency and shown in Fig.~\ref{Fig5}. It is clear that, with a specific photon frequency $\omega_{r}$, $P_{\textrm{LZ}}^{\textrm{L}}$ and $P_{\textrm{LZ}}^{\textrm{R}}$ have the same oscillation frequency, which is independent on the phonon frequency $\omega_{\textrm{ph}}$. In addition, the oscillation frequency is linearly proportional to the photon frequency. It indicates that the oscillation frequency of $P_{\textrm{LZ}}$ is solely dependent on the frequency of the photon mode and equal to $\omega_{r}/\pi$ precisely speaking.

\section{conclusions}
\label{Conclusions}

Employing the multi-D$_2$ {\it Ansatz} to describe the coupled qubit-photon dynamics in a Rabi dimer system containing driven qubits, we explore the modulations of the photon-assisted LZ transitions by tuning the amplitudes and the initial phases of the driving fields. These two parameters of the driving fields determine the energy diagram of a coupled qubit-photon monomer, which inherently dominates the coupled qubit-photon dynamics. It is found that photon-assisted ARPs can be achieved once the driving fields are parameterized to fulfill certain criteria, i.e., the existence of avoided crossings (a large enough driving field amplitude), the adiabaticity of the process (a strong enough qubit-photon coupling or the sufficient number of photons), and a large initial detuning between the neighbouring adiabatic states with opposite qubit polarizations. Therefore, the time points at which ARPs occur and the time stride between two adjacent ARPs can be precisely engineered by tuning the driving field amplitude and initial phase.

Beyond the impacts from the driving field amplitudes and initial phases, the effects of a micromechanical resonator (modelled as a low-frequency phonon mode coupled to the two qubits) on the dynamics of the hybrid QED system are further unveiled within the framework of this study. Only the left qubit is driven by a driving field in this part in order to highlight the phonon effects on the system dynamics. A phonon mode with a lower frequency can be easily populated. However, it needs a longer time to accumulate sufficient energy to affect the qubits. Increasing the qubit-phonon coupling strength $\alpha$ accelerates the energy flow from the driving field through the left qubit to the phonon mode. Thus the phonon effects on the coupled qubit-photon dynamics emerge earlier with a stronger $\alpha$, and can even destroy the photon-assisted ARPs of the left qubit. In contrast to the phonon, which tends to confine the qubits in one state, the photons are off-diagonally coupled to the qubits and apt to flip the qubits. Therefore, the photon effects on the qubit dynamics can be clearly noticed from the high-frequency oscillations of the LZ transition probability $P_{\textrm{LZ}}$ of the qubits. The frequencies of these oscillations are found solely dependent on the photon frequencies. 

It is expected that the knowledge gained in this work on the controllable ARPs with driving fields and phonon mode can benefit the quantum state preparation and engineering in quantum information and quantum computing with QED apparatus.

\section*{Acknowledgments}
Support for this project from the Nanyang Technological University under the Undergraduate Research Experience on CAmpus (URECA) programme and from the Singapore Ministry of Education Academic Research Fund (Grant No.~RG87/20) is gratefully acknowledged. We also thank Yuejun Shen for the helpful discussion. 

\appendix*
\section{The time-dependent variational approach}
\label{Equations of Motion}

The variational principle results in equations of motion of the variational parameters as follows,

\begin{eqnarray}  
&&i\sum_{n=1}^{M}\left[\dot{A}_{n}+A_{n}\Theta_{ln}\right]\bar{S}_{ln}\nonumber\\
&&=\sum_{n=1}^{M} \Bigg\{ A_{n} \Big[ \mathbcal{F}_{\rm L} + \mathbcal{F}_{\rm R} + \Xi_{ln} + 2\lambda(\eta_{l}^{*}+\eta_{n}) \Big] \nonumber\\
&&-g \Big[ C_{n}(\mu_{l}^{*}+\mu_{n}) - B_{n}(\nu_{l}^{*}+\nu_{n}) \Big] \Bigg\} \bar{S}_{ln},
\end{eqnarray}
where 

$\Theta_{ln} = \mu_{l}^{*}\dot{\mu}_{n}+\nu_{l}^{*}\dot{\nu}_{n}+\eta_{n}^{*}\dot{\eta}_{n}$,

$\bar{S}_{ln}=\exp\left[\mu_{l}^{*}\mu_{n}+\nu_{l}^{*}\nu_{n}+\eta_{l}^{*}\eta_{n}\right]$,

$\mathbcal{F}_{\rm L} = F_{\rm L} \cos(\Omega_{\rm L}t + \text{\ensuremath{\Phi_{\rm L}}}) /2$,

$\mathbcal{F}_{\rm R} = F_{\rm R} \cos(\Omega_{\rm R}t + \text{\ensuremath{\Phi_{\rm R}}}) /2$,

$\Xi_{ln} = \omega_{\rm L}\text{\ensuremath{\mu_{l}^{*}}}\mu_{n}+\omega_{\rm R}\text{\ensuremath{\nu_{l}^{*}}}\nu_{n}-J\left(\mu_{l}^{*}\nu_{n}+\mu_{n}\nu_{l}^{*}\right)+\omega_{\textrm{ph}}\eta_{l}^{*}\eta_{n}$.

\begin{eqnarray}   
&&i\sum_{n=1}^{M}\left[\dot{B}_{n}+B_{n}\Theta_{ln}\right]\bar{S}_{ln} = \sum_{n=1}^{M} \Bigg\{ B_{n} ( \mathbcal{F}_{\rm L} - \mathbcal{F}_{\rm R} + \Xi_{ln} ) \nonumber\\
&&-g \Big[ D_{n}\left(\mu_{l}^{*}+\mu_{n}\right) + A_{n}\left(\nu_{l}^{*}+\nu_{n}\right) \Big]\Bigg\} \bar{S}_{ln},
\end{eqnarray}

\begin{eqnarray}   &&i\sum_{n=1}^{M}\left[\dot{C}_{n}+C_{n}\Theta_{ln}\right]\bar{S}_{ln} = \sum_{n=1}^{M} \Bigg\{ C_{n} (-\mathbcal{F}_{\rm L} + \mathbcal{F}_{\rm R} + \Xi_{ln}) \nonumber\\
&&-g \Big[A_{n}\left(\mu_{l}^{*}+\mu_{n}\right) + D_{n}\left(\nu_{l}^{*}+\nu_{n}\right) \Big] \Bigg\} \bar{S}_{ln},
\end{eqnarray}

\begin{eqnarray}   
&&i\sum_{n=1}^{M}\left[\dot{D}_{n}+D_{n}\Theta_{ln}\right]\bar{S}_{ln}\nonumber\\
&&= \sum_{n=1}^{M} \Bigg\{ D_{n} \Big[-\mathbcal{F}_{\rm L} - \mathbcal{F}_{\rm R} + \Xi_{ln} -2\lambda\left(\eta_{l}^{*}+\eta_{n}\right)\Big]\nonumber\\
&&-g \Big[ B_{n}\left(\mu_{l}^{*}+\mu_{n}\right) + C_{n}\left(\nu_{l}^{*}+\nu_{n}\right) \Big] \Bigg\} \bar{S}_{ln}.
\end{eqnarray}

\begin{align}
\begin{autobreak}
\MoveEqLeft
i\sum_{n=1}^{M} \left[\left(A_{l}^{*}\dot{A}_{n}+B_{l}^{*}\dot{B}_{n}+C_{l}^{*}\dot{C_{n}}+D_{l}^{*}\dot{D}_{n}\right)\mu_{n}\right.
+\left(A_{l}^{*}A_{n}+B_{l}^{*}B_{n}+C_{l}^{*}C_{n}+D_{l}^{*}D_{n}\right)\dot{\mu_{n}}
+\left(A_{l}^{*}A_{n}+B_{l}^{*}B_{n}+C_{l}^{*}C_{n}+D_{l}^{*}D_{n}\right)
\left.\times\left(\mu_{l}^{*}\dot{\mu}_{n}+\nu_{l}^{*}\dot{\nu}_{n}+\eta_{n}^{*}\dot{\eta}_{n}\right)\mu_{n}\right]\bar{S}_{ln}
=i\sum_{n=1}^{M} \frac{\mu_{n}}{2}\left[\left(A_{l}^{*}A_{n}+B_{l}^{*}B_{n}-C_{l}^{*}C_{n}-D_{l}^{*}D_{n}\right) \times \mathbcal{F}_{\rm L} \right]\bar{S}_{ln}
+i\sum_{n=1}^{M} \frac{\mu_{n}}{2} \left[\left(A_{l}^{*}A_{n}-B_{l}^{*}B_{n}+C_{l}^{*}C_{n}-D_{l}^{*}D_{n}\right) \times \mathbcal{F}_{\rm R} \right]\bar{S}_{ln}
+\sum_{n=1}^{M}\mu_{n}\bar{S}_{ln}\left(A_{l}^{*}A_{n}+B_{l}^{*}B_{n}+C_{l}^{*}C_{n}+D_{l}^{*}D_{n}\right)
\times(\omega_{\rm L}\text{\ensuremath{\mu_{l}^{*}}}\mu_{n}+\omega_{\rm R}\text{\ensuremath{\nu_{l}^{*}}}\nu_{n}-J\left(\mu_{l}^{*}\nu_{n}+\mu_{n}\nu_{l}^{*}\right)+\omega_{\textrm{ph}}\eta_{l}^{*}\eta_{n})
+\sum_{n=1}^{M}\bar{S}_{ln}\left(A_{l}^{*}A_{n}+B_{l}^{*}B_{n}+C_{l}^{*}C_{n}+D_{l}^{*}D_{n}\right)
\times\left(\omega_{\rm L}\mu_{n}-J\nu_{n}\right)
-g\sum_{n=1}^{M}\mu_{n}\left(A_{l}^{*}C_{n}+B_{l}^{*}D_{n}+C_{l}^{*}A_{n}+D_{l}^{*}B_{n}\right)
\times\left(\mu_{l}^{*}+\mu_{n}\right)\bar{S}_{ln}
-g\sum_{n=1}^{M}\mu_{n}\left(A_{l}^{*}C_{n}+B_{l}^{*}D_{n}+C_{l}^{*}A_{n}+D_{l}^{*}B_{n}\right)\bar{S}_{ln}
-g\sum_{n=1}^{M}\mu_{n}\left(A_{l}^{*}B_{n}+B_{l}^{*}A_{n}+C_{l}^{*}D_{n}+D_{l}^{*}C_{n}\right)
\times\left(\nu_{l}^{*}+\nu_{n}\right)\bar{S}_{ln}
+\sum_{n=1}^{M}\mu_{n}\left(2A_{l}^{*}A_{n}-D_{l}^{*}D_{n}\right)\lambda\left(\eta_{l}^{*}+\eta_{n}\right)\bar{S}_{ln},
\end{autobreak}
\end{align}
\begin{align}
\begin{autobreak}
\MoveEqLeft
i\sum_{n=1}^{M} \left[\left(A_{l}^{*}\dot{A}_{n}+B_{l}^{*}\dot{B}_{n}+C_{l}^{*}\dot{C_{n}}+D_{l}^{*}\dot{D}_{n}\right)\nu_{n}\right.
+\left(A_{l}^{*}A_{n}+B_{l}^{*}B_{n}+C_{l}^{*}C_{n}+D_{l}^{*}D_{n}\right)\dot{\nu_{n}}
+\left(A_{l}^{*}A_{n}+B_{l}^{*}B_{n}+C_{l}^{*}C_{n}+D_{l}^{*}D_{n}\right)
\left.\times\left(\mu_{l}^{*}\dot{\mu}_{n}+\nu_{l}^{*}\dot{\nu}_{n}+\eta_{n}^{*}\dot{\eta}_{n}\right)\nu_{n}\right]\bar{S}_{ln}
=i\sum_{n=1}^{M} \nu_{n}\left[\left(A_{l}^{*}A_{n}+B_{l}^{*}B_{n}-C_{l}^{*}C_{n}-D_{l}^{*}D_{n}\right) \times \mathbcal{F}_{\rm L} \right]\bar{S}_{ln}
+i\sum_{n=1}^{M} \nu_{n} \left[\left(A_{l}^{*}A_{n}-B_{l}^{*}B_{n}+C_{l}^{*}C_{n}-D_{l}^{*}D_{n}\right) \times \mathbcal{F}_{\rm R} \right]\bar{S}_{ln}
+\sum_{n=1}^{M}\nu_{n}\bar{S}_{ln}\left(A_{l}^{*}A_{n}+B_{l}^{*}B_{n}+C_{l}^{*}C_{n}+D_{l}^{*}D_{n}\right)
\times(\omega_{\rm L}\text{\ensuremath{\mu_{l}^{*}}}\mu_{n}+\omega_{\rm R}\text{\ensuremath{\nu_{l}^{*}}}\nu_{n}-J\left(\mu_{l}^{*}\nu_{n}+\mu_{n}\nu_{l}^{*}\right)+\omega_{\textrm{ph}}\eta_{l}^{*}\eta_{n})
+\sum_{n=1}^{M}\bar{S}_{ln}\left(A_{l}^{*}A_{n}+B_{l}^{*}B_{n}+C_{l}^{*}C_{n}+D_{l}^{*}D_{n}\right)
\times\left(\omega_{\rm R}\nu_{n}-J\mu_{n}\right)
-g\sum_{n=1}^{M}\nu_{n}\left(A_{l}^{*}C_{n}+B_{l}^{*}D_{n}+C_{l}^{*}A_{n}+D_{l}^{*}B_{n}\right)
\times\left(\mu_{l}^{*}+\mu_{n}\right)\bar{S}_{ln}
-g\sum_{n=1}^{M}\nu_{n}\left(A_{l}^{*}C_{n}+B_{l}^{*}D_{n}+C_{l}^{*}A_{n}+D_{l}^{*}B_{n}\right)\bar{S}_{ln}
-g\sum_{n=1}^{M}\nu_{n}\left(A_{l}^{*}B_{n}+B_{l}^{*}A_{n}+C_{l}^{*}D_{n}+D_{l}^{*}C_{n}\right)
\times\left(\nu_{l}^{*}+\nu_{n}\right)\bar{S}_{ln}
+\sum_{n=1}^{M}\nu_{n}\left(2A_{l}^{*}A_{n}-D_{l}^{*}D_{n}\right)\lambda\left(\eta_{l}^{*}+\eta_{n}\right)\bar{S}_{ln},
\end{autobreak}
\end{align}

and
\begin{align}
\begin{autobreak}
\MoveEqLeft 
i\sum_{n=1}^{M}\left[\left(A_{l}^{*}\dot{A}_{n}+B_{l}^{*}\dot{B}_{n}+C_{l}^{*}\dot{C_{n}}+D_{l}^{*}\dot{D}_{n}\right)\eta_{n}\right.
+\left(A_{l}^{*}A_{n}+B_{l}^{*}B_{n}+C_{l}^{*}C_{n}+D_{l}^{*}D_{n}\right)\dot{\eta}{}_{n}
+\left(A_{l}^{*}A_{n}+B_{l}^{*}B_{n}+C_{l}^{*}C_{n}+D_{l}^{*}D_{n}\right)
\left.\times\left(\mu_{l}^{*}\dot{\mu}_{n}+\nu_{l}^{*}\dot{\nu}_{n}+\eta_{n}^{*}\dot{\eta}_{n}\right)\eta_{n}\right]\bar{S}_{ln}
= 
i\sum_{n=1}^{M}\eta_{n}\left[\left(A_{l}^{*}A_{n}+B_{l}^{*}B_{n}-C_{l}^{*}C_{n}-D_{l}^{*}D_{n}\right) \times \mathbcal{F}_{\rm L} \right]\bar{S}_{ln}
+i\sum_{n=1}^{M} \eta_{n} \left[\left(A_{l}^{*}A_{n}-B_{l}^{*}B_{n}+C_{l}^{*}C_{n}-D_{l}^{*}D_{n}\right) \times \mathbcal{F}_{\rm R} \right]\bar{S}_{ln}
+\sum_{n=1}^{M}\eta_{n}\bar{S}_{ln}\left(A_{l}^{*}A_{n}+B_{l}^{*}B_{n}+C_{l}^{*}C_{n}+D_{l}^{*}D_{n}\right)
\times(\omega_{\rm L}\text{\ensuremath{\mu_{l}^{*}}}\mu_{n}+\omega_{\rm R}\text{\ensuremath{\nu_{l}^{*}}}\nu_{n}-J\left(\mu_{l}^{*}\nu_{n}+\mu_{n}\nu_{l}^{*}\right)+\omega_{\textrm{ph}}\eta_{l}^{*}\eta_{n})
+\sum_{n=1}^{M}\left(A_{l}^{*}A_{n}+B_{l}^{*}B_{n}+C_{l}^{*}C_{n}+D_{l}^{*}D_{n}\right)\omega_{\textrm{ph}}\eta_{n}\bar{S}_{ln}
-g\sum_{n=1}^{M}\eta_{n}\left(A_{l}^{*}C_{n}+B_{l}^{*}D_{n}+C_{l}^{*}A_{n}+D_{l}^{*}B_{n}\right)
\times\left(\mu_{l}^{*}+\mu_{n}\right)\bar{S}_{ln}  
-g\sum_{n=1}^{M}\eta_{n}\left(A_{l}^{*}B_{n}+B_{l}^{*}A_{n}+C_{l}^{*}D_{n}+D_{l}^{*}C_{n}\right)
\times\left(\nu_{l}^{*}+\nu_{n}\right)\bar{S}_{ln}
+\sum_{n=1}^{M}\eta_{n}\left(2A_{l}^{*}A_{n}-2D_{l}^{*}D_{n}\right)\lambda\left(\eta_{l}^{*}+\eta_{n}\right)\bar{S}_{ln}
+\sum_{n=1}^{M}\left(2A_{l}^{*}A_{n}-D_{l}^{*}D_{n}\right)\lambda\bar{S}_{ln}.
\end{autobreak}
\end{align}

By numerically solving these linear equations at each time $t$, one can calculate the values of $\dot{A}_{n}$, $\dot{B}_{n}$, $\dot{C}_{n}$, $\dot{D}_{n}$, $\dot{\mu}_{n}$, $\dot{\nu}_{n}$, and $\dot{\eta}_{n}$ accurately. The fourth-order Runge-Kutta method is then adopted for the time evolution of the tunable Rabi dimer, including the time-dependent photon numbers, phonon number, qubit polarization, LZ transition probability.

\end{document}